# An Unsupervised Machine Learning-based Framework for Wafer Scale Variability Analysis and Performance Prediction of Ferroelectric $Hf_{0.5}Zr_{0.5}O_2$ Thin Film Capacitors

*Anika Anu, Sayani Majumdar**
*Faculty of Information Technology and Communication Sciences, Tampere University, Finland*
**Email: sayani.majumdar@tuni.fi*

## Abstract

Fabrication process-induced performance variability remains a formidable barrier in the high-volume manufacturing of semiconductor chips. With skyrocketing Artificial Intelligence (AI) workload, demand for non-volatile and computational memories is growing exponentially. As embedded non-volatile memory, ferroelectric $Hf_{0.5}Zr_{0.5}O_2$ emerged as a strong candidate due to their CMOS back-end-of-line (BEOL) compatibility, scalability and high performance. However, their sensitive crystallization kinetics leads to significant device-to-device (D2D) non-uniformity leading to unpredictability of performance over wafer scale. In this work, we demonstrate unsupervised machine learning can analyze intra-die D2D variations and predict performance of "unseen" dies efficiently. We present a framework utilizing Principal Component Analysis (PCA) and K-Means clustering to analyze D2D performance variations in HZO capacitors and building on data from multiple dies, we move beyond traditional descriptive statistics to a predictive "Virtual Metrology" approach that separates performance categories, defined by key parameters like remanent polarization ($P_r$) and coercive voltage ($V_c$). The analysis further extends to comparing uniformity across different dies across the wafer showing the proposed methodology can accurately predict device performance on untested dies with a low Mean Absolute Percentage

Error (MAPE) in the range of 5 - 10%, suggesting a robust path for accelerated yield improvement and reduced metrology overhead.

## Introduction

Artificial intelligence (AI) has become a part of every automation in today's world. For semiconductor manufacturing, it has been supporting different tasks for instance in inspection tools, statistical process control, yield analysis and so on [1-3]. However, with emergence of more and more complex device fabrication techniques, role of AI needs to evolve from their traditional image classifier approaches to more advanced design partners.

In the AI era, advanced memory research became one of the major exploration direction. Device architectures are becoming more complex to meet the needs of memory and compute at the same location and process windows getting tightened due to needs from heterogeneous and 3D vertical integration process limitations [4-6]. Reliability of device performance within these tight process windows are a major concern. To consistently predict wafer-scale performance variations, a large characterization dataset is often required. Inclusion of Machine Learning (ML) can help in effectively finding a pattern from small number of dataset and predict performances of uncharacterized dies and can significantly improve timeline for manufacturing in semiconductor fabs, accelerating the development timeline.

For AI and edge computing workload, non-volatile embedded memory and near or in-memory computing is one of the most active research direction. As an energy-efficient, fast, memory solution with possibilities of multibit data retention over long time durations, ferroelectric devices have emerged as a promising solution [7,8]. Ferroelectric random-access memories (FeRAM) are the most matured technology among the group that are already available in the market while the ferroelectric field effect transistors (FeFETs) are becoming the next big players as one-transistor memory solution. In both architectures, a ferroelectric capacitor (FeCap) acts an essential component for storing the charges in terms of voltage-induced ferroelectric polarization switching. Both FeFETs and FeCaps are considered promising candidates for resistive and capacitive in-memory computing (IMC) for next generation of AI hardware [9,10]. Due to 3-terminal architectures of FeRAM and FeFETs, area overhead is a concern for IMC circuits. To address this challenge 2-terminal ferroelectric tunnel junctions (FTJ) are also under intense study where resistive IMC is possible within a compact area overhead [11,12]. For all these architectures, development of ultrathin, conformal ferroelectric layer is essential.

In the last decade, ferroelectric hafnium zirconium oxide, $Hf_{0.5}Zr_{0.5}O_2$ (HZO) has emerged as one of the most promising materials for the next wave of advanced semiconductor memory devices, showing high potential for integration in embedded non-volatile memory and neuromorphic circuits [13,14]. Large polarization switching ($P_r$) in 10 nm or less ultrathin ferroelectric HZO layer obtained within low process temperatures [15], high endurance [16], non-volatile retention of polarization [17] allows their integration in FeRAMs, FeFETs or FTJ structures where it is possible to open a large memory window and maintain it over extended period of time without external power supply. These properties make them very attractive for edge-inference devices where low-energy operation is mandatory to sustain operation with limited battery power [18-20].

The ferroelectricity in $HfO_2$ based Fluorites is primarily attributed to the stabilization of the non-centrosymmetric orthorhombic phase ($Pca2_1$), a state influenced by factors such as electrode and capping-layer-induced strain, doping concentration, thermal annealing and so on [18,21]. Among these Flourites, HZO is particularly attractive due to its lower crystallization temperature compared to other dopants, making it highly compatible with CMOS back-end-of-line (BEOL) processing [19]. However, recent research emphasizes that the stoichiometry and interface engineering between the HZO and the electrodes are critical, as oxygen vacancy formation and interfacial layer formation during crystallization annealing significantly impact the polarization switching kinetics and phase purity [20]. Furthermore, understanding the spatial distribution of these defects is essential to understand the inherent device-to-device (D2D) performance variability that causes reliability concerns in large-scale memory arrays [22].

As previously stated, the key advantage of HZO lies in the robust ferroelectricity down to ultrathin film thickness, even down to 1 nm and compatibility with the standard CMOS manufacturing process [23]. However, despite these advantages, the full potential of HZO-based systems is not realized currently due to significant reliability challenges in the low thermal budget HZO. Long term reliability issues such as time dependent dielectric breakdown (TDDB), fatigue and endurance over prolonged bias stress are one of the major concerns [16,24,25] while short term reliability issues like performance inhomogeneity over wafer scale is another. Therefore, it is important to fully understand the physical mechanisms behind these reliability issues, including cycle-to-cycle (C2C) and D2D variation, fatigue under continuous bias stress resulting in imprint and eventual device failure. These are critical research challenges as this insight is essential for designing high endurance, reliable hardware [16]. Also, understanding performance variations among different dies within the same wafer is essential to improve process parameters and yield. All of these require large amounts of data which is extremely resource hungry and require high level of human expertise. For the semiconductor industry, this is a big challenge that needs automation.

Considering ferroelectric HZO case, the inconsistent device behavior shows up in terms of wide variety of values of remanent polarization ($P_r$) and coercive voltage ($V_C$), that defines the most critical parameter like memory window. The variations often arise from incomplete crystallization of HZO at BEOL compatible processes with low annealing temperature [17,26]. To understand the inhomogeneous performance problems in general and to predict the dies with reliable performance, semiconductor fabs rely on wafer-scale characterization. As mentioned previously, the main challenge for that is having extensive measurements for mapping process-induced variability across a full wafer. Additionally, probing and measuring thousands of devices across a wafer creates a massive data bottleneck that slows down research and development cycles and delays process optimization. This is where traditional test-and-measure methods fall short.

ML-based techniques can reduce the measurement workload by identifying correlations in a smaller representative subset of data and automate the performance prediction task which can shorten the development cycle. Instead of just being a tool to analyze massive datasets after they have been collected, ML can be used to understand variations and predict performance much faster and from much less data [27]. This approach has already shown significant promise for reliability engineering in these specific devices [28].

Deep Learning techniques like Convolutional Neural Networks (CNNs) are widely used for wafer-scale defect identification and analysis due to their strong performance in image classification tasks [29-31]. Since wafer maps can be treated as images, CNNs enable end-to-end detection of faulty dies and learn without manual feature engineering and have shown promising results across many studies. [32,33] Recent advances in this direction include Error-correcting CNN models [34], hybrid approaches such as CNN together with handcrafted features, Quantum–classical hybrid methods and techniques for mixed-type defect patterns [35]. However, deep learning-based techniques come with the challenges like need for large number of parameters resulting in slow training and inference and for dynamic manufacturing environments, frequent retraining is needed. Also, for high classification accuracy, the model requires large, high-quality labeled datasets. As labeling is manual, costly, and time-consuming, the training becomes expensive and quality of dataset depends on expert consistency and protocol, introducing some more variabilities there.

Therefore, it is essential to build a model that learns the complex, underlying patterns from a small, representative set of measured devices by identifying some characteristic features in the dataset. This model can then forecast the performance distribution across an entire wafer, drastically reducing the need for exhaustive, time-consuming measurements.

In this work, we develop a novel framework where we use unsupervised machine learning specifically Principal Component Analysis (PCA) and K-Means clustering method to analyze variability in HZO devices, fabricated partially and characterized in our own laboratories. We show that PCA can untangle complex, multi-parameter data to find the few key factors that drive performance. We then use clustering techniques to automatically group devices with similar performances. Finally, we build a predictive model based on this analysis, showing that we can forecast device properties with Mean Absolute Percentage Error (MAPE) between 5 – 10 % for $P_r$ and between 1.53% and 5.74% for $V_c$ from a limited device characterization dataset, paving the way for faster, more intelligent process characterization and yield improvement.

## 2. Methods

In this section, we first define the main concept for this work (**Figure 1**). The detailed process flow for this work is given in Supplementary information **Figure S1**. Next, we describe individual processes like device fabrication and characterization results and finally utilization of this data in developing the framework.

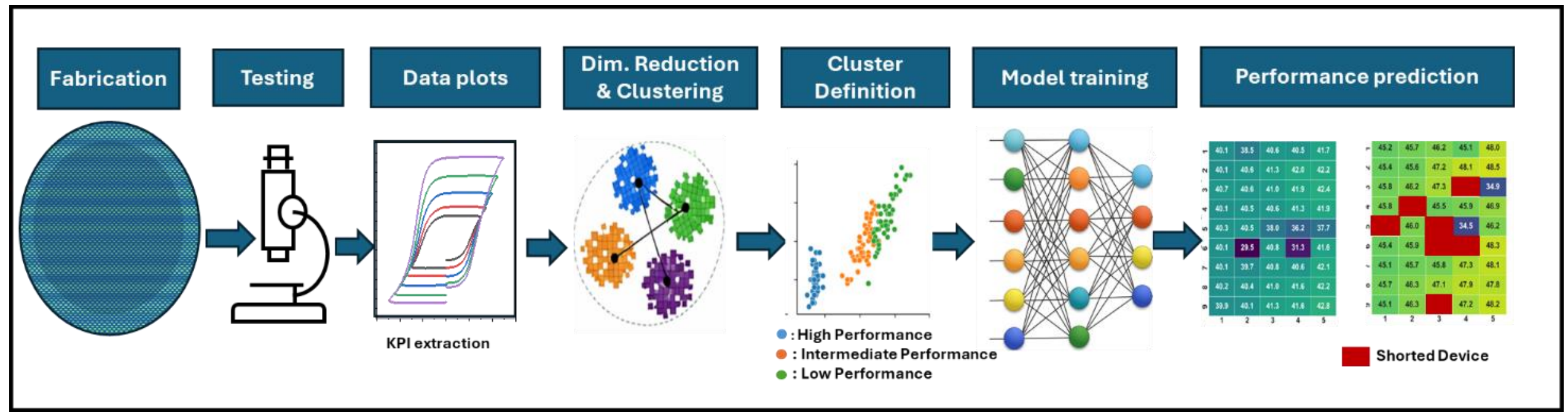

**Figure 1:** Schematic diagram of the workflow starting from ferroelectric HZO-based capacitor structure fabrication, their electrical characterization, data plotting and key performance indices (KPI) extraction followed by development of machine learning (ML) based virtual metrology framework for ferroelectric device performance prediction for non-characterized dies. The framework consists of dimensionality reduction using principal component analysis (PCA) and clustering, principal component definition (PCD), model training and performance prediction.

## 2.1 Device Fabrication

The devices analyzed in this study were fabricated on a cluster Atomic Layer Deposition (ALD) tool [26]. The ferroelectric capacitor stack consisted of a 30 nm TiN bottom electrode (grown at 300 °C), a 10 nm HZO ferroelectric layer (grown at 200 °C, both using thermal ALD), and a protective $Al_2O_3$ cap layer of 2 nm. The process was conducted in a Applied™ Picosun™ single wafer cluster ALD system, without a vacuum break after the TiN growth to ensure a clean interface between the metal and ferroelectric layer. Following deposition, the samples underwent a Post-Deposition Anneal (PDA) via Rapid Thermal Annealing (RTA) at 600 °C for 30 seconds to crystallize the HZO film into the ferroelectric orthorhombic phase. Finally, Ti (2 nm)/Au (50 nm) layers were formed for top electrode metallization. Detailed structural properties of the films are discussed in our previous work [26]. Electrical measurements were performed across multiple dies on the wafer.

## 2.2 Electrical Characterization

Electrical characterization of the samples was carried out by applying voltage between the bottom TiN and the top gold electrode. For ferroelectric polarization vs. voltage (*P*-*V*) loops, Ferroelectric Material Tester (Aixacct TF analyzer 2000) was used. To keep the analysis uniform, Ferroelectric loops measured on 200 × 200 μm$^2$ square junctions are only used in this analysis. $P$-$V$ loops are measured using dynamic hysteresis measurement (DHM) techniques with a triangular voltage sweep of 1 kHz. From the $P$-$V$ loops, the remanent polarization ($P_r$) - the polarization remaining when the electric field is zero and the coercive voltage ($V_c$) - the voltage required to switch the polarization were extracted. A total of 270 devices were measured across six distinct die locations in a quarter-wafer sample, resulting in a comprehensive dataset of intra-die and inter-die variations. Measurements were taken from multiple devices across several dies to build a comprehensive dataset of intra-die and inter-die variations. All measurements were carried out at room temperature and under ambient conditions.

## 2.3 Machine Learning Analysis

The data analysis was performed using a Python script with the scikit-learn library [27]. The workflow consisted of two main stages:

1. **Unsupervised Analysis:** In the first stage Principal Component Analysis (PCA) was used for dimensionality reduction to visualize the measured $P – V$ dataset and identify the main sources of variation. Then K-Means clustering technique was used to automatically group devices into performance categories, with the optimal cluster count ($k$) determined by the "elbow method."

   The K-Means algorithm partitions the ferroelectric parameter dataset into $k$ clusters by minimizing the within-cluster Sum of Squared Errors (SSE), or inertia, defined as:

$$SSE = \sum_{j=1}^{k} \sum_{x \in C_j} \left|\left|x - \mu_j\right|\right|^2$$

   where $C_j$ is the $j^{th}$ cluster and $\mu_j$ is the centroid of that cluster [36]. Each device is assigned to a performance category based on the proximity rule $Cluster(x) = \arg\min_j \left|\left|x - \mu_j\right|\right|^2$.

2. **Predictive Modeling:** A predictive model was built based on the unsupervised clustering. The model was trained on data from a set of measured dies as the "training" dies. Its purpose is to predict the performance of the cluster and key electrical parameters ($P_r$, $V_c$) of devices on separate, "uncharacterized" dies. The model's accuracy was then verified by comparing its predictions against the actual experimental data from that uncharacterized die, with performance quantified by the Mean Absolute Percentage Error (MAPE).

The MAPE is used to quantify the accuracy of the "Virtual Metrology" forecast by measuring the average relative displacement between predicted and experimental values:

$$MAPE = \frac{100\%}{n} \sum_{i=1}^{n} \left| \frac{A_i - P_i}{A_i} \right|$$

Where n is the number of devices, $A_i$ is the actual measured value ($P_r$ or $V_c$), and $P_i$ is the predicted parameter [37].

## 3. Results and Discussion

### 3.1. Ferroelectric properties of HZO FeCaps

To verify the basic electrical behavior of our fabricated HZO capacitors **Figure 2(a**), we examined their polarization vs. voltage (*P-V*) loops using dynamic hysteresis measurement (DHM). Voltage was applied on the top Au electrode keeping the bottom TiN electrode grounded. Physical location of the measured dies on one-quarter of a 200 mm wafer is shown in **Figure 2(b).** The map shows the measured dies, labeled as Die 1, Die 2. Die 3, Die 4, Die 5, and Die 6, were first predicted using the ML-framework and then experimentally measured to verify the prediction and also to train the model on more experimental data. **Figure 2(c)** shows the *P-V* characteristics of the devices of two different dies named Die 1, marked in green and Die 2, marked in red in **Figure 2(b)**. The measurement results of the two dies bring our main challenge into focus. The *P-V* loop measurements from Die 2 show all data from a single die are similar with very little D2D variation while devices from Die 1 show a noticeable spread in both remanent polarization ($P_r$) and coercive voltage ($V_C$). This variation suggests that although the devices consistently exhibit ferroelectricity, device-to-device variations, an issue commonly reported in HZO systems [24,38] remain substantial. This performance variability affects the wafer-scale yield significantly since devices with properties far from a standard value cannot be utilized in functional circuits. To improve the yield and bring down cost of fabrication, therefore we need to better understand performance range of different dies over the wafer.

### 3.2. Device-to-Device (D2D) Variability Analysis

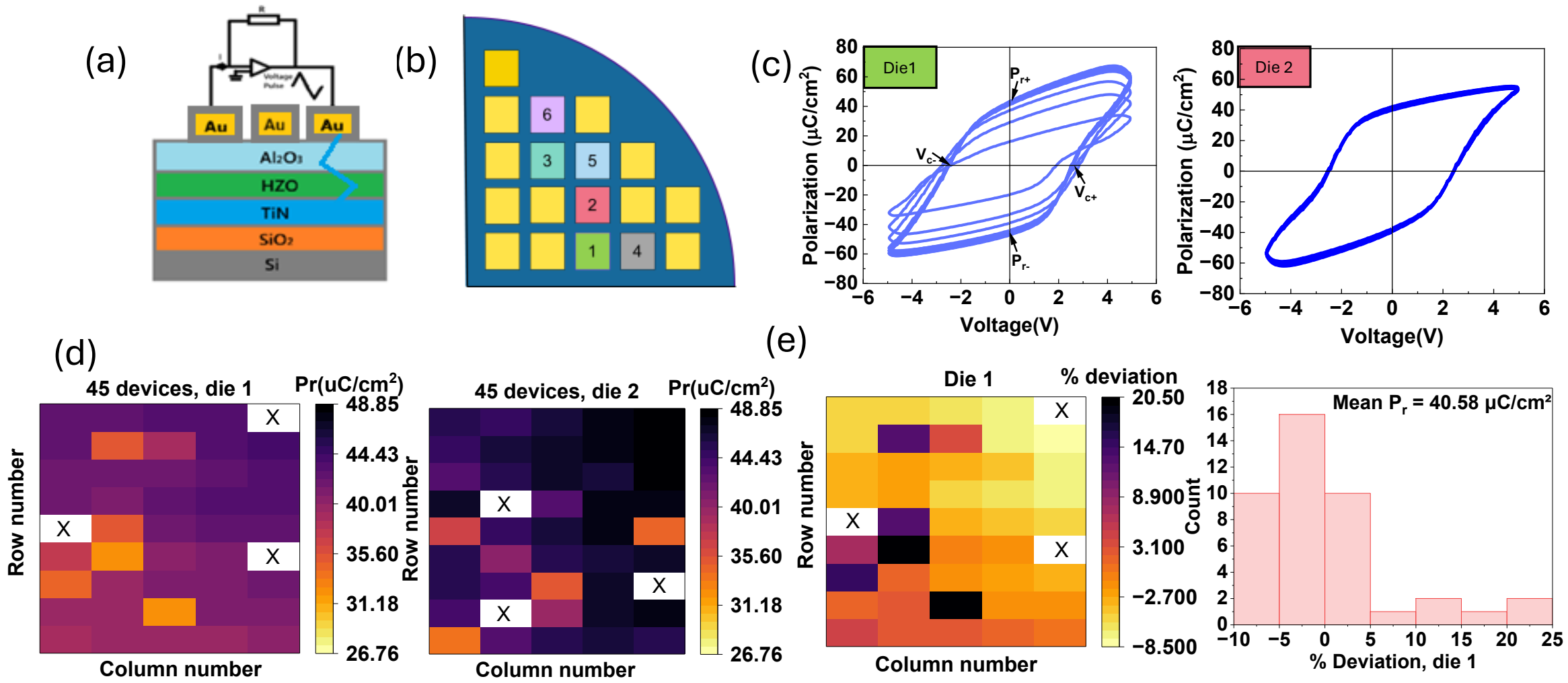

**Figure 2:** **(a)** Schematic diagram of the HZO capacitor device structure used in this study together with electrical contacts used for measurement of the capacitors. **(b)** Position on the wafer for the die 1, die 2, die 3, die 4, die 5 and die 6. **(c)** Measured *P-V* loops from multiple devices on die 1 and die 2. **(d)** Distribution of $P_r$ values among devices within die 1 and die 2 (Cross signs represents shorted device). **(e)** Deviation from mean $P_r$ value for single devices and bar plot showing number of devices with % variation from the mean value for die 1.

Local performance mapping in **Figure 2(d)** revealed inconsistent intra-die homogeneity across the wafer. Analysis of the *P-V* hysteresis data showed a high functional yield exceeding 97% for both measured dies, with the statistical distribution across working devices demonstrating low parametric variability, typically less than 10% device-to-device variation in key metrices with a few outliers. However, this local consistency failed to translate to wafer-scale uniformity, highlighting a significant challenge in understanding wafer-scale yield. When performance signatures were compared across different dies, such as Die 1 and Die 2, a clear inter-die parametric drift was observed. These dies exhibited distinct statistical profiles characterized by differing median $P_r$ values and increased distribution spreads. This complex multi-scale variability ranging from tight local grouping to broad global drift is exceedingly difficult to analyze and control using conventional statistical methods thus validating the necessity of a machine learning (ML) driven framework.

### 3.3. Introduction to Machine Learning Framework

The electrical data shown in the previous section has high dimensionality, strong variable correlation and multi-scale non-uniformity that motivates the need for advanced ML techniques for analysis and prediction. The application of AI/ML in semiconductor manufacturing is rapidly growing, focusing on process optimization, defect detection and accelerated metrology [39]. For large-scale data analysis and prediction various models have been employed in materials science, for example supervised learning where models like Random Forests or Artificial Neural Networks (ANNs) are often used for direct prediction from labeled data, establishing predictive relationships between input factors (e.g., process parameters, deposition rates) and output performance (e.g., predicted $P_r$ value) [39]. In unsupervised learning techniques like principal component analysis (PCA) and K-Means clustering are ideal for identifying hidden, intrinsic patterns within high-dimensional device data, often revealing root causes of variation without prior labeling [33].

Our approach utilizes this unsupervised methodology to first reduce the complexity of the data and reveal the intrinsic clusters of device behavior. We then leverage this clustering to build our predictive framework, aiming to significantly reduce the expensive, time-consuming measurement steps required for full wafer characterization.

### 3.4. Unsupervised Clustering of Device Performance

Initial analysis focused on quantifying the D2D variability of key ferroelectric parameters. Box plots of average remanent polarization ($P_r$) as a function of hysteresis amplitude revealed that the largest performance variation occurred at a measurement voltage of 5V, as shown in **Figure 3(a)**. This voltage was therefore selected for subsequent in-depth analysis. To prepare for clustering, the elbow method was applied to the dataset to determine the optimal number of device clusters ($k$) [27]. The plot of the sum of squared errors (SSE) versus the number of clusters showed a distinct "elbow" at $k$=3, as shown in **Figure 3(b)**, indicating that the devices naturally group into three primary categories based on their electrical characteristics.

As shown in the PCA plot of intra-die clusters (**Figure 3(c)**), the $k$ = 3 clustering model successfully separates the device population into three distinct groups. By analyzing the contributions of the original electrical parameters to the principal components, we can assign physical meaning to the axes. The analysis revealed that Principal Component 1 (PC1) represents as a 'Performance Axis,' primarily correlated with the magnitude of $P_r$, while Principal Component 2 (PC2) functions as an 'Asymmetry Axis,' capturing shifts in the positive and negative coercive voltages ($V_c$). The three clusters were identified as Cluster 0: High Performance, Cluster 1: Intermediate Performance, and Cluster 2: Low Performance (as confirmed by the data values in each group).

Further comparison between two separate dies using this framework showed that Die 2 has significantly more uniformity (**Figure 3(d)**) with its device performance clustering more tightly. In contrast, die 1 exhibited higher overall $P_r$ values but with greater device-to-device spread. This methodology provides a clear and intuitive way to visualize and quantify process uniformity across a wafer. **Figure 3(e)** shows the distributions of $P_r$ values at 5V applied voltage across multiple measured dies.

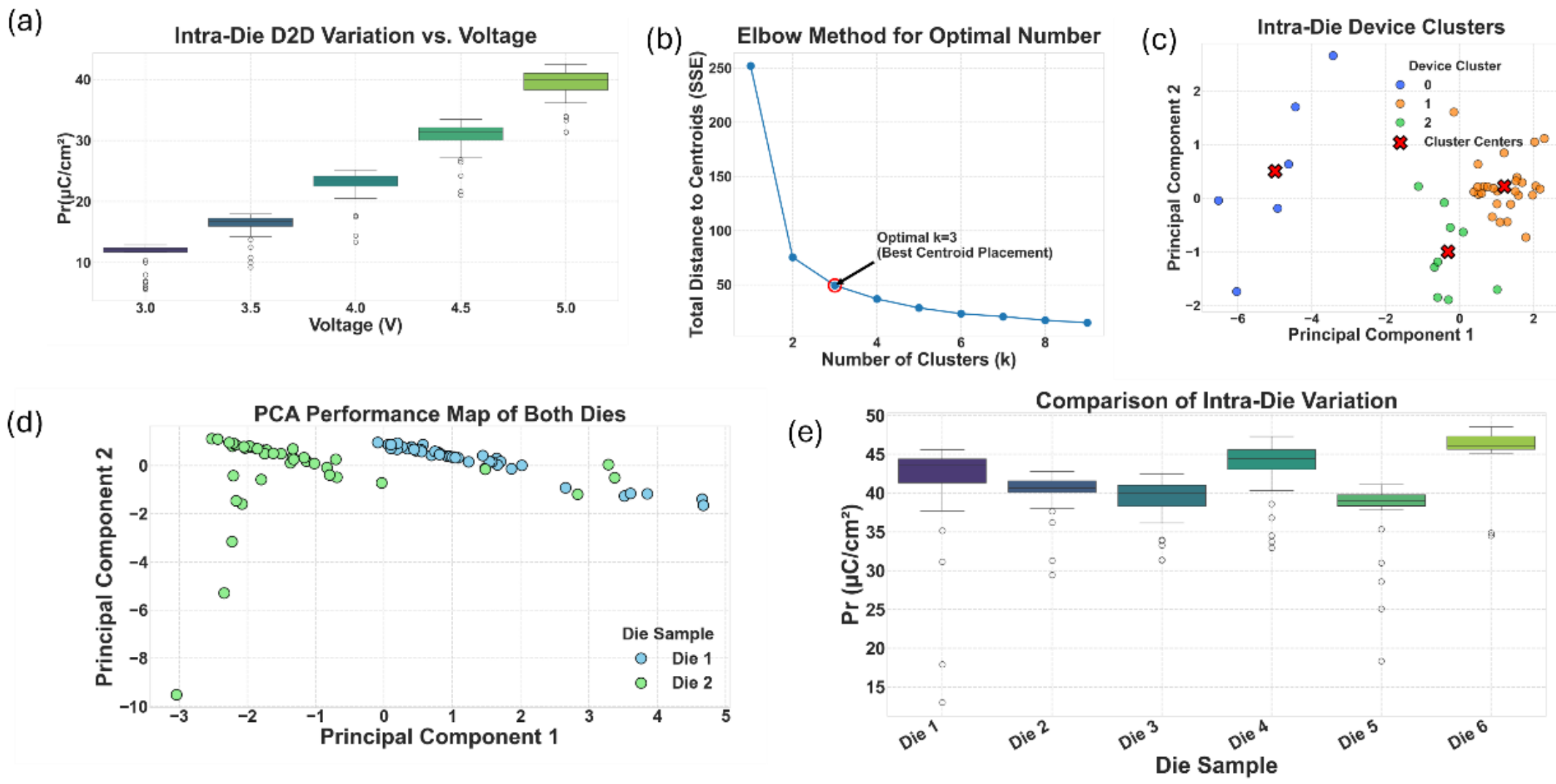

**Figure 3**: Analysis of optimal clustering and variation visualization **(a)** Intra-Die D2D Variation at 5V. **(b)** Elbow method for optimal numbers, SSE is calculated based on the distance to the centroids. **(c)** Intra-Die device clusters at 5V(k=3). **(d)** PCA performance map of both dies. **(e)** Comparison of intra-Die variation of $P_r$ values measured at 5V for different dies.

Once the dimensionality reduction, clustering and performance definition is successfully done, we proceed with testing whether the devices on the unseen dies would lie close to the defined performance clusters with the predicted placement of devices on the PCA map (stars) and closely match the actual measured data (circles). This quantitative verification is shown in **Figure 4**.

### 3.5. Predictive Modeling and Verification

The final and most critical step was to move from variability analysis to prediction. We trained our model on the cluster centers and data from several "training" dies and then tasked it with predicting the performance of entirely unseen dies. A prediction was made for the properties of the new dies like $P_r$ and $V_C$.

The model was trained iteratively starting with data from two measured dies and progressively increasing the training dataset to four and then five dies. The model's ability to capture the full range of variation improved as data from more dies were included in the training showing that the model is better able to capture the true wider range of performance variability across the wafer when trained on wider range of datasets. For instance, with a limited training dataset of just two dies, the prediction accuracy for the unseen die, Die 3 achieved a MAPE for $P_r$ of 5.92%. However, with training on only two-die's dataset, the prediction for unknown dies might not be a representative value. Therefore, initially, the training set is expanded to four dies. In this case, the MAPE for $P_r$ for the unseen Die 5 was 10.91%, showing if performance variation between dies is bigger, the MAPE can increase, even after being trained on more experimental data. So, we further expanded the dataset and found that with five dies of training data, the MAPE for $P_r$ for Die 6 was 7.87%. The data is given in **Table 1**. This trend demonstrates that after incorporating 4-5 die's training data, the framework can achieve robust performance prediction and MAPE does not exceed ~10% value, highlighting its practical utility for reducing metrology overhead.

**Table 1.** Training data on different dies vs. MAPE for $P_r$ and $V_C$ values.

| Prediction of Dies and their training dataset | MAPE $P_r$ (%) | MAPE $V_C$+ (%) | MAPE $V_C$- (%) |
|---|---|---|---|
| Die 3 - After training with data from die 1, 2 | 5.92 | 1.53 | 2.95 |
| Die 5 - After training with data from die 1, 2, 3, 4 | 10.91 | 1.65 | 2.54 |
| Die 6 - After training with data from die 1, 2, 3, 4, 5 | 7.87 | 5.74 | 5.73 |

In **Figure 4** we quantify the model's accuracy by comparing the actual measured values of $P_r$ and $V_C$ against the predicted values for each device clusters across the three unseen dies. The "Actual vs. Predicted" plots show a remarkable correlation. For Die 3 in **Figure 4(a)**, the predicted cluster 0 lie very close to the cluster center leading to a MAPE scores of 5.92% for $P_r$, 1.53% for $V_C$+, and 2.95% for $V_{C\text{-}}$. Similarly, in Figure 4(b), the predicted Clusters 1 and 2 for Die 5 are positioned near their respective centroids, despite the higher inter-die drift, the model maintains a MAPE of 10.91% for $P_r$, 1.65% for $V_C$+, and 2.54% for $V_C$-. In Figure 4(c), the predicted Cluster 1 for Die 6 resides in close proximity to the established cluster center, achieving MAPE scores of 7.87% for $P_r$, 5.74% for $V_C$+ and 5.73% for $V_C$-.This ensures that all predicted data points are more or less close to the "Ideal Prediction" line across all three dies, demonstrating that the model is not just sorting data it has already seen but can forecast the average performance of unmeasured devices with high accuracy, at least within the process space explored in this study. The consistent alignment of these "unseen" device signatures with pre-defined cluster centers visually validates the transition from standard descriptive statistics to a functional "Virtual Metrology" approach.

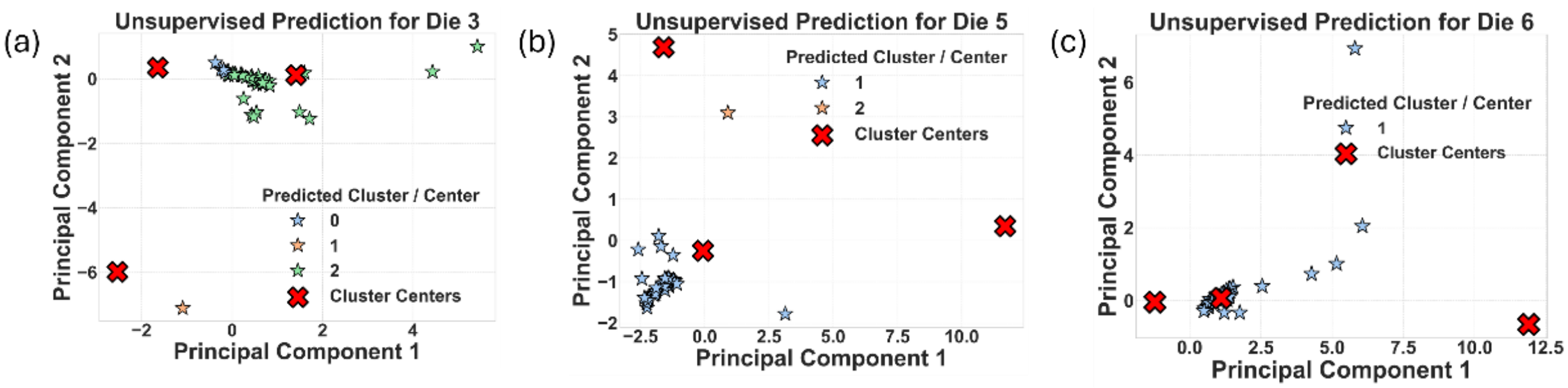

**Figure 4**: Verification of the predictive model showing how close the cluster center lies to different predicted device clusters. Results are from three "unseen" dies, **(a)** Die 3 **(b)** Die 5 **(c)** Die 6.

To evaluate the efficacy of the unsupervised machine learning framework in capturing wafer-scale variability, we performed a probabilistic comparison between experimental measurements and model forecasts using Gaussian distributions, as shown in **Figures 5 (a-c)**. The Experimental Gaussian Fit (represented by the solid blue line) was generated by applying a Kernel Density Estimation (KDE) to the ground truth data extracted from the "unseen" dies. This method utilizes a Gaussian kernel to transform discrete electrical parameters, specifically the average $P_r$, $V_{C+}$ and $V_{C-}$ into a continuous probability density function, providing a clear visual representation of the D2D variability.

The Predicted Gaussian Fit (represented by the dashed red line) was derived through a two-step computational process. First, each device on the uncharacterized die was mapped to its most probable performance category using the K-Means clustering model. The model assigned the mean parameter value of the corresponding cluster center to each device, effectively creating a "Virtual Metrology" forecast. To account for the expected statistical dispersion within a predicted performance group and to prevent zero variance artifacts in the visualization, a synthetic Gaussian noise (jitter) was introduced to the predicted values. This noise was scaled to 15% of the experimental standard deviation, ensuring that the predicted distribution reflects the expected physical spread of a realistic device population.

The statistical distributions are modeled using a Gaussian Probability Density Function (PDF) [40]:

$$f(x) = \frac{1}{\sigma\sqrt{2\pi}} e^{-\frac{1}{2}\left(\frac{x-\mu}{\sigma}\right)^2}$$

To ensure the model reflects physical D2D variation, the predicted values $P_{jitter}$ incorporate a synthetic noise component: $P_{jitter} = \mu_{cluster} + \epsilon$, where $\epsilon \sim \mathcal{N}\left(0, \left(0.15\sigma_{exp}\right)^2\right)$

The strong spatial and statistical correlation between the experimental and predicted distributions serves as a robust validation of the model's accuracy. As indicated by the low MAPE scores reaching as low as 5.92% for $P_r$ in Die 3 the model successfully bridges the gap between discrete cluster-based classification and continuous performance forecasting. This alignment demonstrates that the framework can accurately characterize the performance landscape of untested dies, significantly reducing the requirement for exhaustive electrical measurements while maintaining high predictive fidelity.

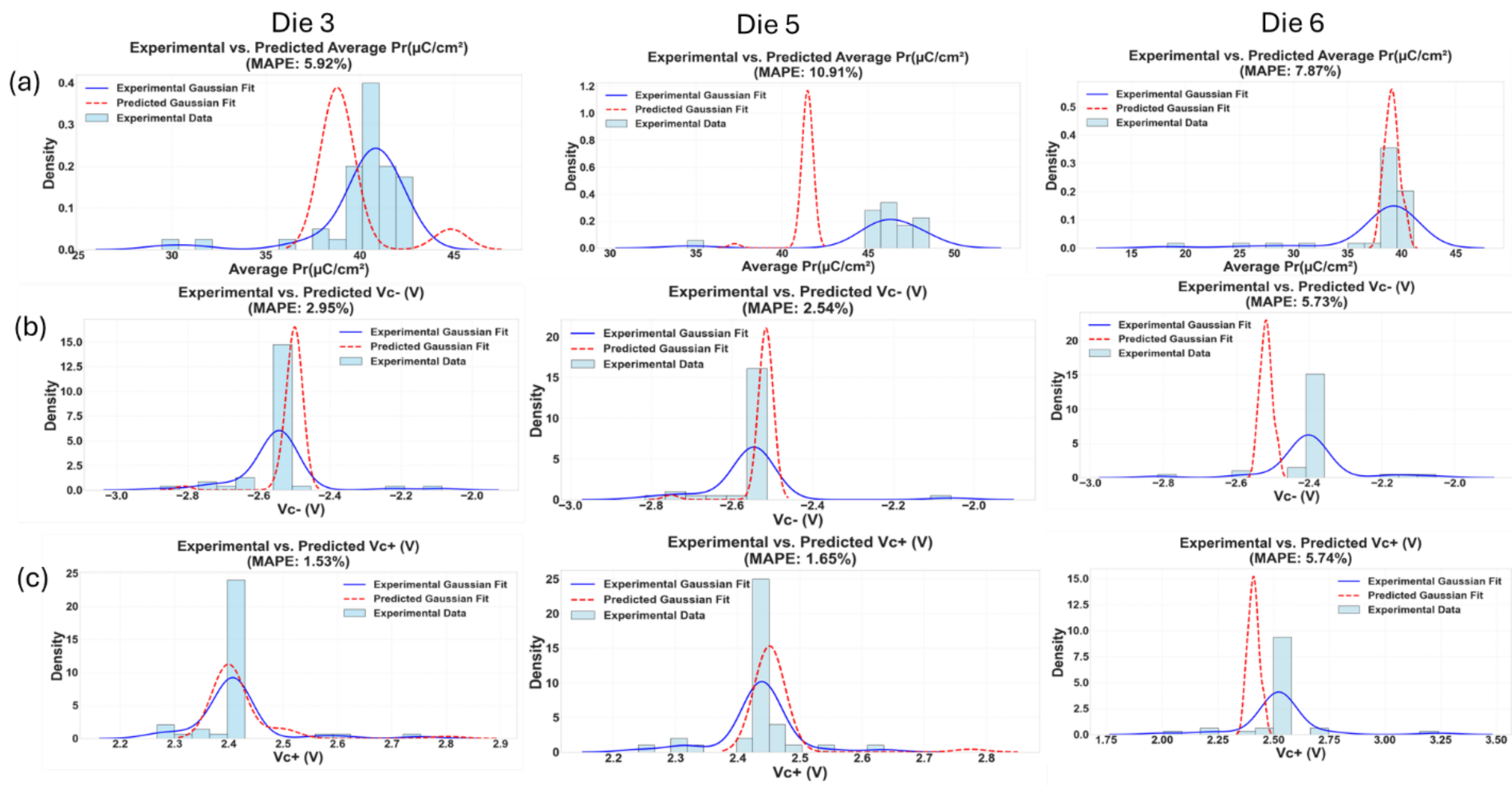

**Figure 5**: Comparison of Experimental and Predicted distribution plots for **(a)** Remanent Polarization ($P_r$), **(b)** Negative Coercive Voltages $V_{C-}$ and **(c)** Positive Coercive Voltages $V_{C+}$ on three dies: Die 3, Die 5 and Die 6. The blue bar plots and blue solid line is the statistical distribution from the measured devices while the dotted red curve presents the predicted Gaussian curve.

In **Figure 5(a),** Die 6 $P_r$ appears to be in better alignment between the predicted and experimental peaks, although there is a higher MAPE (7.87%) than for Die 3. This discrepancy is attributed to MAPE's inherent sensitivity to low magnitude actual values. The experimental outliers in the 15-30 μC/cm$^2$ range in Die 6 results high relative errors that disproportionately inflate the average. In comparison, the data in Die 3 is more tightly clustered, despite the slight mean shift, the absence of low value outliers ensures a lower overall MAPE (5.92%).

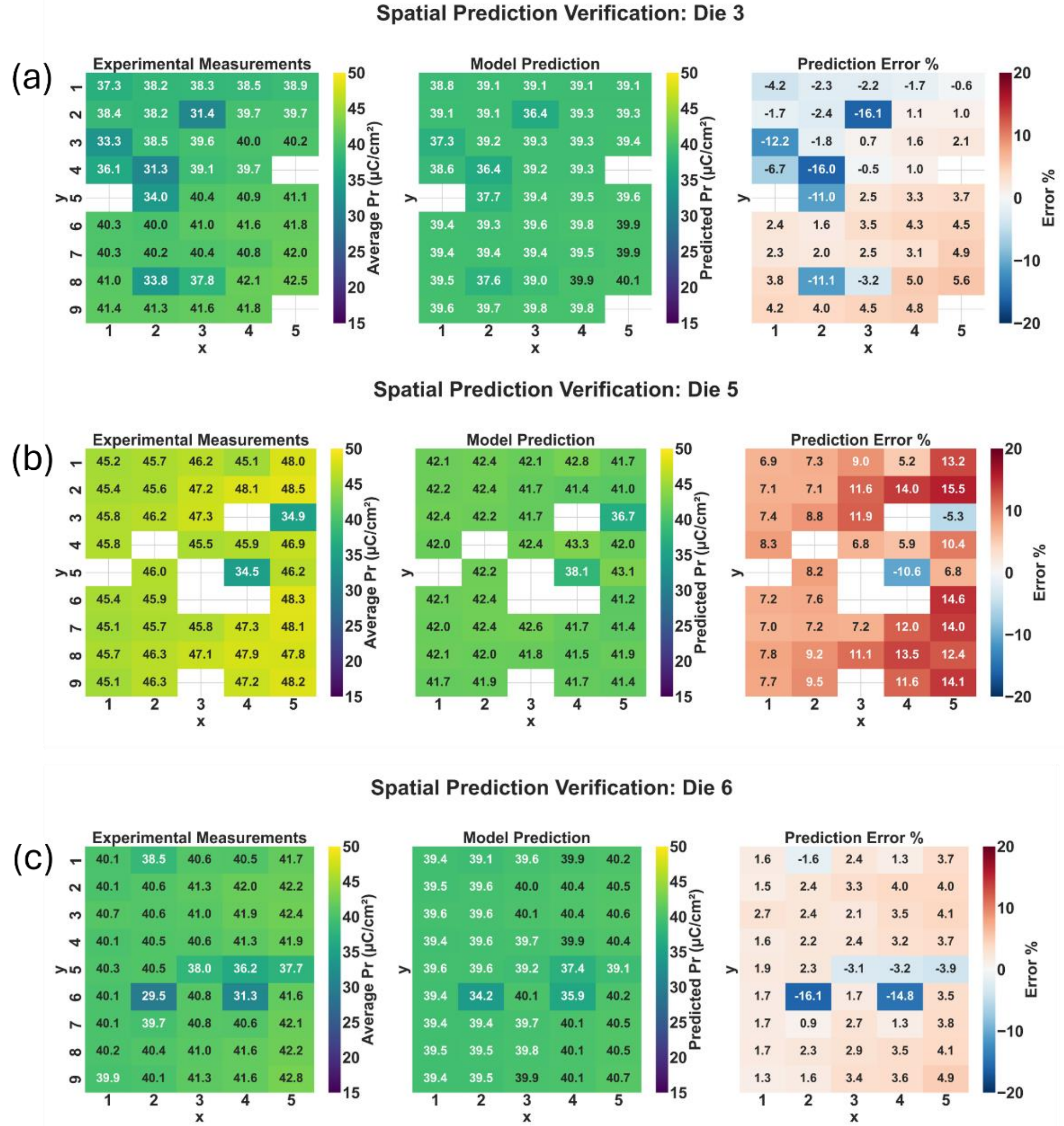

**Figure 6:** Spatial Prediction Verification of Remanent Polarization ($P_r$) values for devices on (a) Die 3, (b) Die 5 and (c) Die 6. The left panel shows experimentally measured $P_r$ numbers from 9 rows and 5 columns of similar devices, the middle panel shows the numbers predicted by the framework and the right column shows the % prediction error for each individual devices.

The final validation step involves assessing the spatial prediction accuracy of single devices across the dies. **Figure 6** presents heatmaps comparing the actual measured $P_r$ distribution from single devices of an unseen die against the model's predicted $P_r$ distribution (derived from cluster centers) and the resulting prediction error percentage. For all three tested dies (Die 3, Die 5, and Die 6), the model accurately captures the overall spatial trends in performance, confirming that the cluster approach effectively links device electrical signatures to their physical location on the wafer.

### 3.6 Universality of the proposed framework

Finally, to understand the universality of the proposed framework, we applied the methods on a different sample, a wafer of HZO capacitors fabricated in the same way as the discussed sample, however annealed at a much lower temperature of 450 °C. This sample has in general much lower $P_r$ values arising from lower crystallization occurring at lower temperatures, however with higher uniformity compared to the higher temperature annealed sample (**Figure S2**). We found that with experimentally measured data from 2 dies, the model can provide prediction for the 3$^{rd}$ die with 9.65% MAPE for $P_r$ and for $V_C$s, the numbers are 3.4% and 0.8% respectively, proving for different wafer level variations, the unsupervised learning model can work with similar accuracy.

To do a comparison with other approaches, a regression-based strategy was also explored for die level prediction. A multiple polynomial regression model was utilized, employing device (x, y) spatial coordinates to predict $P_r$ as a continuous surface for the spatial prediction verification. which is simpler to implement, computationally lightweight, and easy to interpret. However, it assumes a smooth spatial function and therefore fails to capture localized $P_r$ anomalies in the data, leading to higher prediction errors at critical positions. The unsupervised learning framework avoids this limitation by learning the variability structure directly from the data, achieving a less than 10% MAPE.

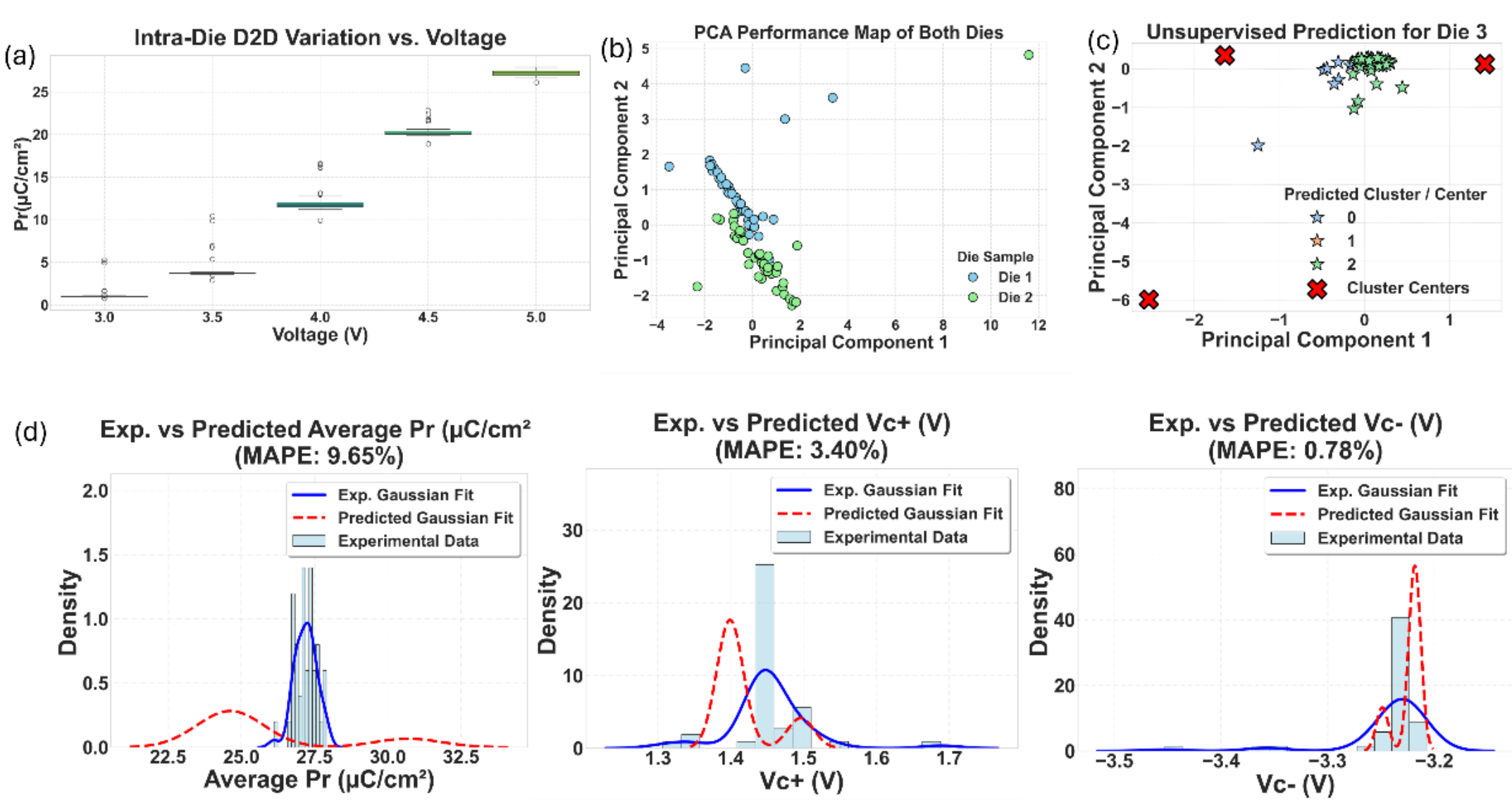

**Figure 7:** Analysis of optimal clustering and variation visualization **(a)** Intra-Die D2D Variation at 5V. **(b)** PCA performance Map of both dies. **(c)** Intra-Die device clusters at 5V (k=3). (d) Experimental vs. Predicted verification plots for Remanent Polarization ($P_r$), Negative Coercive Voltages $V_c$- and Positive Coercive Voltages Vc+ on two dies.

### 3.7 Physics-based understanding of the results

In order to understand the cause for the performance variations and to have better predictions beyond the purely mathematical approach, it is important to have a deeper look at the device physics and the factors that can modify device performance over wafer-scale. As discussed before, ferroelectricity in HZO evolves from orthorhombic phase (*o*- phase) formation during post deposition annealing of the capacitor stack. Now different parameters play a crucial role in determining the *o*- phase, including temperature of deposition [24,41] annealing temperature [14, 25], strain induced by electrodes [18] presence of capping layer, oxygen vacancy formation [19, 20] and so on. For uniform strain on the HZO structure over wafer scale, it is important to have uniform thickness of bottom TiN electrode layer. Also, thermal distribution on the wafer during rapid thermal annealing step could play a significant role. Some hot spots and some colder spots on the wafer not only can change crystallization of HZO on the particular spots but can lead to higher oxygen vacancy formation in the hot spots since higher temperature increases the oxidation of TiN electrodes and HZO gets more oxygen vacancies losing oxygen in those areas. Higher concentration of oxygen vacancies increases interface and bulk trap states impacting $P_r$ and leakage currents that on the long-term affects the bias-stress-related reliability performances like endurance and TDDB.

One observable change in the D2D variability data is observed when samples were grown at a higher temperature of 280 ºC [24] or annealed for longer period in a standard furnace which can be attributed to more uniform heat distribution on the wafer [42]. It is found that larger grain formation due to high temperature growth plays a major role in reducing D2D variation [24]. Strain distribution, chemical analysis and grain size mapping over the wafer scale are under study currently and in future this data, combined with the electrical data will provide a more complete picture and help in development of Physics-informed ML model-based prediction framework.

## 3. Conclusion

In conclusion, a combined hardware-machine-learning based framework to access chip-scale and wafer-scale variability of ferroelectric HZO capacitors for ferroelectric random-access memories is presented in this work for the first time. It is found that unsupervised machine learning techniques, specifically Principal Component Analysis (PCA) and K-Means clustering, can effectively categorize complex device-to-device non-uniformity, however, with the initial requirement of characterizing a representative training subset to map the multi-parameter variations. By leveraging these identified performance clusters, a predictive model is developed to estimate device performance on unmeasured dies using limited training data with its peak accuracy dependent on progressively capturing the wider range of process variability across the wafer. Both spatial trends and continuous performance forecasts were successfully validated showing mean absolute percentage errors (MAPE) approximately 10% for key parameters on unseen dies, including a highly accurate 5.92% for remanent polarization ($P_r$). Our results show that transitioning from traditional exhaustive test and measure methods to this data-driven "Virtual Metrology" approach can lead to faster process characterization, significantly reduced metrology overhead, and a robust path for accelerated yield improvement.

Supplementary information

# An Unsupervised Machine Learning-based Framework for Wafer Scale Variability Analysis and Performance Prediction of Ferroelectric $Hf_{0.5}Zr_{0.5}O_2$ Thin Film Capacitors

*Anika Anu, Sayani Majumdar**

*Faculty of Information Technology and Communication Sciences, Tampere University, Finland*

**Email: sayani.majumdar@tuni.fi*

## S1. Methodological workflow

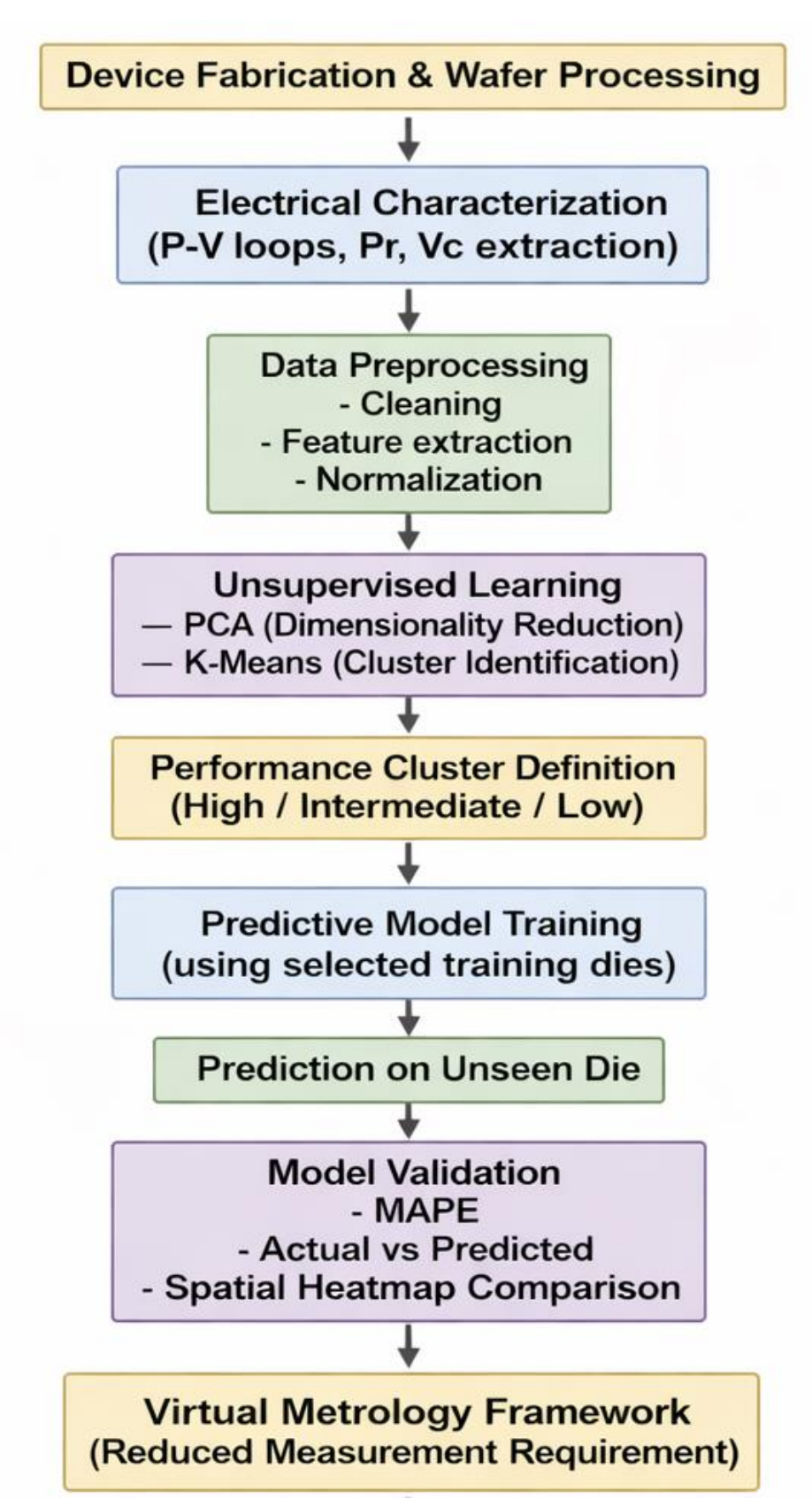

**Figure S1:** Overview of a data-driven virtual metrology workflow for device characterization and prediction.

## S2. Wafer scale performance mapping from experiments

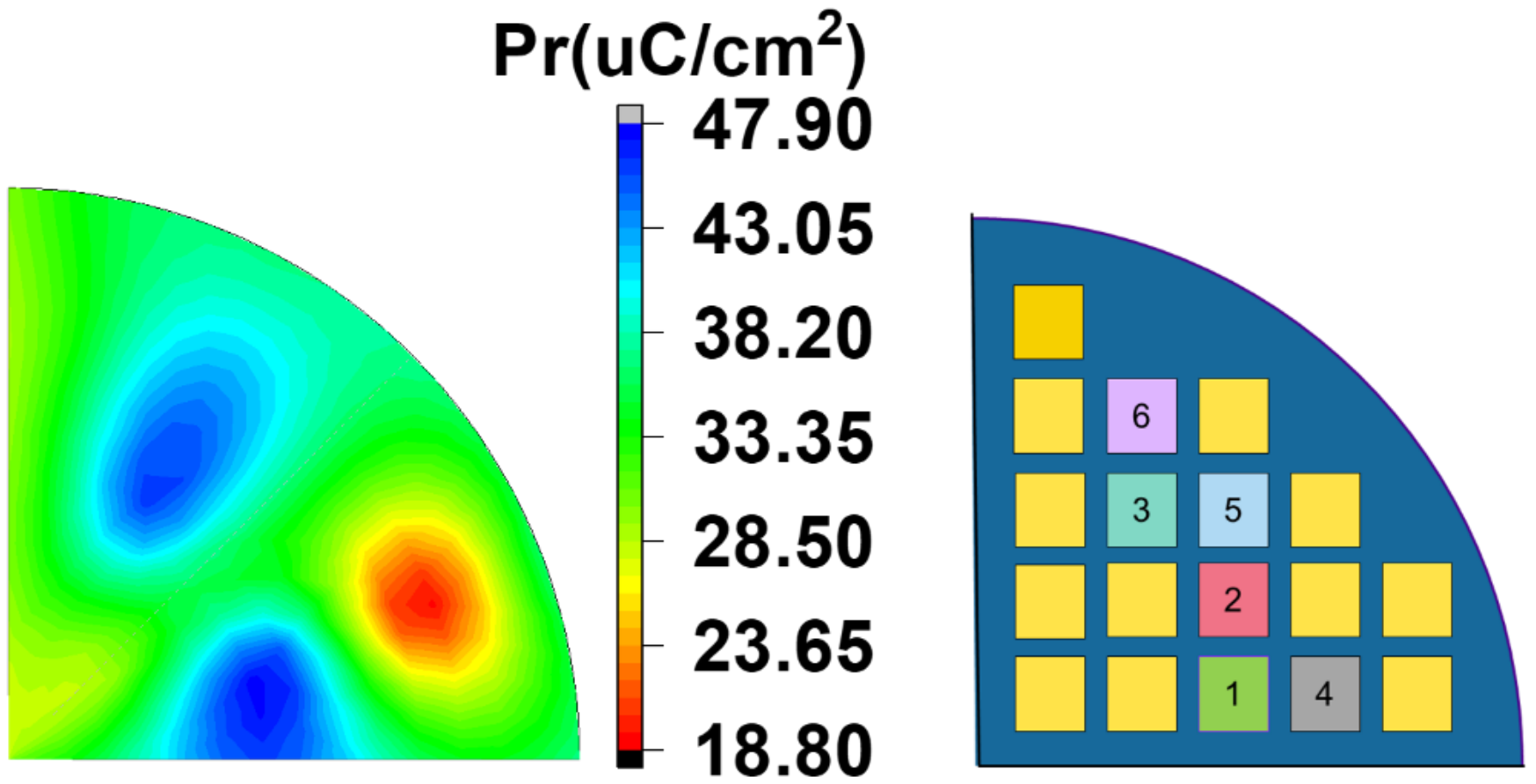

**Fig. S2:** Mapping of the remanent polarization ($P_r$) values from the measured devices from one-quarter of a wafer (on the left) and die location of on the wafer (right) for the sample annealed at 600 °C showing there are physical locations on the wafer where $P_r$ values are significantly different from the mean $P_r$ value. Prediction on dies in those regions make the error percentage higher.

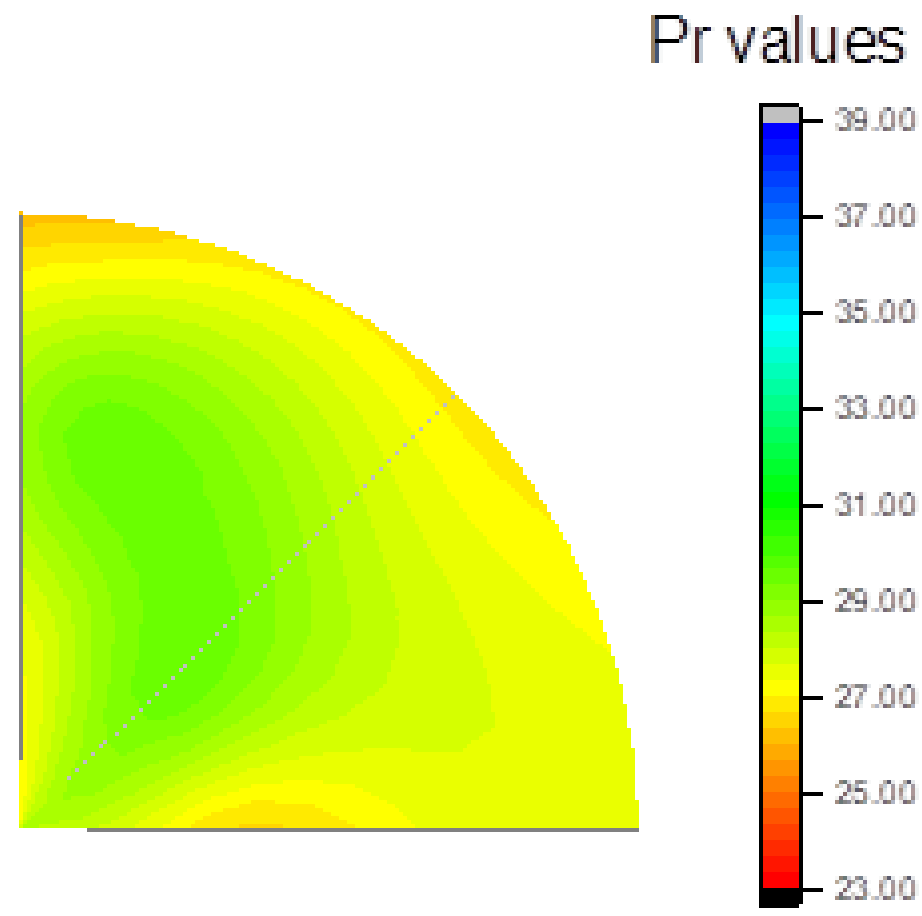

**Fig. S3:** Mapping of the remanent polarization ($P_r$) values from the measured devices from one-quarter of a wafer for the sample annealed at 450 °C showing the spatial nonuniformity comes mainly from the rapid thermal annealing as both 600 °C (Fig. S2) and 450 °C annealed samples are part of the same 200 mm wafer.